\overfullrule 0mm

 \input harvmac.tex
\voffset 1cm
\input amssym.def
\input amssym.tex
\input epsf.tex

\def\slh{\widehat{sl}}

\def\CA{{\cal A}}

\def\CV{{\cal V}}\def\CV{{\cal V}}

\def\bro{\bar \rho}

\def\bw{\overline w}

\def\tW{\tilde{W}}
\def\bW{\overline W}

\def\tn{\bar{n}}
%\overline{n}}
%\def\tq{\tilde q}

\def\hN{\hat{N}}
\def\za{\alpha} \def\zb{\beta}
\def\zg{\gamma} 
\def\zl{\lambda}\def\zs{\sigma}
\def\zL{\Lambda}
\def\hchi{\hat{\chi}}
\def\Gl{\lambda}\def\GL{\Lambda}\def\Gr{\rho}
\def\Gm{\mu}\def\Gn{\nu}

\def\Go{\omega}\def\zo{\omega}

\def\CG{{\cal G}}
\def\CA{{\cal A}}\def\CP{{\cal P}}
\def\h{h}
%\kappa}

\def\rep{representation}
\font\maju=cmtcsc10
\def\nimrep{{{\maju nim}}-rep}

\def\Exp{{\rm Exp}\,}
\def\mod{{\rm\  mod\,}}  \def\omit#1{{}}
\def\nind{\par\noindent}
%%%%%%%%%%%%%%%
\def\plb#1#2#3{Phys. Lett. {\bf B #1} (#2) #3}
\def\npb#1#2#3#4{Nucl. Phys. {\bf B #1} [FS#2] (#3) #4}
\def\cmp#1#2#3{Comm. Math. Phys. {\bf  #1} (#2) #3}

\def\hepth#1{{\tt hep-th/#1}}
\def\mathph#1{{\tt math-ph/#1}}

\def\mathOA#1{{\tt math.OA/#1}}
\def\mathQA#1{{\tt math.QA/#1}} 
 %%%%%%%%%%%%%%%%%%%%%%%%%%

\lref\DIFZ{P. Di Francesco and J.-B. Zuber,
% $SU(N)$ lattice integrable models associated with graphs, 
 \npb{338}{}{1990}{602-646}.}

\lref\BPPZ{R.E. Behrend, P.A. Pearce, V.B. Petkova and J.-B.
Zuber, \plb{444}{1998}{163-166},
\hepth{9809097};
%{On the classification of bulk and boundary conformal field theories};
\npb{579}{}{2000}{707-773}, \hepth{9908036}.}  

\lref\Pasq{V. Pasquier, 
% Two-dimensional critical systems labelled by Dynkin diagrams, 
\npb{285}{}{1987}{162-172}.}

\lref\Xu{F. Xu,
%{\it  New braided endomorphisms from conformal inclusions}
%{\em Comm. Math. Phys.} 192 : 349-403,  1998.        
\cmp{192}{1998}{349-403}.}

\lref\AO{ A. Ocneanu, {\it The classification of subgroups of quantum
$SU(N)$}, 
Lectures at Bariloche Summer School, Argentina, Jan 2000, to appear in
 {\it AMS Contemporary Mathematics}, R. Coquereaux, A. Garcia and
R. Trinchero, eds.}

\lref\Z{J.-B. Zuber, CFT,BCFT ADE and all that, \hepth{0006151},
  Lectures at Bariloche 
Summer School, Argentina, Jan 2000, to appear in
 {\it AMS Contemporary Mathematics}, R. Coquereaux, A. Garcia and
R. Trinchero, eds.}

\lref\Qu{T. Quella, {\it Branching rules of semi-simple Lie algebras 
using affine extensions}, \mathph{0111020};
A. Alekseev, S. Fredenhagen, T. Quella and V. Schomerus, in preparation;
 C. Schweigert, unpublished.}

\lref\SF{C. Schweigert and J. Fuchs, PRHEP-tmr2000/..., TMR
conference ``Non-perturbative quantum effects 2000'', Paris, September
2000, \hepth{0007174}.}

\lref\FRS{J. Fuchs, I. Runkel and C. Schweigert, 
{\it Conformal boundary conditions and 3D topological field theory }, 
\hepth{0110158}.}

\lref\KO{A.N. Kirillov and V. Ostrik, {\it On q-analog of McKay
correspondence and ADE classification of  $\widehat{sl}_2$
 conformal field theories}, \mathph{0101219}; %\semi
V. Ostrik, {\it Module categories, weak Hopf algebras and modular
invariants },
\mathQA{0111139}.}

\lref\BEK{J. B\"ockenhauer and D.E.  Evans,
%  Modular invariants, graphs and $\alpha$-induction for nets of subfactors,
\cmp{200}{1999}{57-103}, \hepth{9805023};
%III : {\em Comm. Math. Phys.}  
\cmp{205}{1999}{183-228},\hepth{9812110}; %\semi
J. B\"ockenhauer, D. E. Evans and Y. Kawahigashi,
\cmp{210}{2000}{733-784},  \mathOA{9907149}.}

\lref\KWFGP{
 V.G. Kac, {\it Infinite-dimensional Lie Algebras}, third
        edition, (Cambridge University Press, 1990); %\semi
%\lref\MW{
M. Walton , 
%       Fusion rules in Wess-Zumino-Witten models, 
%           Nucl. Phys. {\bf B340}, 777--790 (1990).}
                   \npb{340}{}{1990}{777-790}; %\semi
%\lref\FGP{
P. Furlan, A.Ch. Ganchev  and V.B. Petkova, 
%             Quantum groups and fusion rule multiplicities,
\npb{343}{}{1990}{205-227}.}
%              Nucl. Phys. {\bf B343}, 205--227 (1990).}

\lref\BFS{L. Birke, J. Fuchs and C. Schweigert, 
Adv.Theor.Math.Phys. 3 (1999) 671-726, \hepth{9905038}.}

\lref\GG{M. Gaberdiel, T. Gannon, {\it Boundary states for WZW models},
\hepth{0202067}.}
                          
%%%%%%%%%%%%%%%%%%%%%%%%%%%%%%%%%%%%%%%%%%%%%%%%%%%%%%%%%%%%%%%%%%%%%%

\font\Huge=cmbx10 scaled \magstep2

\font\Hugit=cmti10 scaled \magstep2
%%%%%%%%%%%%%%%%%%%%%%%%%%%%%%%
%\draftmode    

%\vbox{\vglue-8mm\baselineskip12pt\hbox{...}\hbox{...} }

\bigskip\centerline{{\Huge Boundary conditions in charge conjugate}}
\centerline{{\Huge {\Hugit sl(N)} WZW theories }}
\bigskip    
\centerline{V.B. Petkova\footnote{${}^\star$}{ E-mail: {petkova@inrne.bas.bg}}}
\centerline{Institute for Nuclear Research and Nuclear Energy}
\centerline{72 Tsarigradsko Chaussee,
1784 Sofia, Bulgaria.}
\centerline{and}
\centerline{J.-B. Zuber\footnote{${}^\dagger$}{ E-mail:  {zuber@spht.saclay.cea.fr}}}
\centerline{SPhT, CEA Saclay, 91191 Gif-sur-Yvette, France.
}

 %%%%%%%%%%%%%%% %%%%%%%%%%%%%%% %%%%%%%%%%%%%%% %%%%%%%%%%%%%%% %%%%%%%%%%%%
\bigskip

\noindent
{\baselineskip9pt {\ninepoint   We compute the representations
(``\nimrep s'') of the fusion algebra of $\slh(N)$ which
determine the boundary conditions of $\slh(N)$ WZW theories
twisted  by the charge conjugation. This is done following two
procedures, one of general validity, the other specific to the
problem at hand.  The problem is related to the classical problem
of decomposition of the fundamental \rep s of $sl(N)$  onto
representations of $B_l=so(2l+1)$ or $C_l =sp(2l)$ algebras.  The
relevant \nimrep s and their diagonalisation matrix are thus
expressed in terms of modular data of the affine $B$ or $C$
algebras. }}

%%%%%%%%%%%%%%%%%%%%%%%%%%%%%%%%%%%%%%%%%%%%%%%%%%%%%%%%%%%%%%%%%%%%%%%%%%%%%%

\newsec{Introduction, notations and results}
\nind
It is now well understood that the possible boundary conditions
of a rational conformal field theory are determined by the set of
non-negative integer valued matrix \rep s, or \nimrep s, of the
fusion algebra of this theory \BPPZ. In the present paper we
address the problem of determining \nimrep s and related data for
those theories of WZW type, that are described by a modular
invariant partition function twisted by charge conjugation
\eqn\modinv{Z=\sum_{\Gl} \chi_\Gl(\tau,z) \chi_{\Gl^*}(\tau,z)^*\ ,
}
To be specific, we restrict here to the $\slh(N)$ current
algebra. This exercise has the double merit of illustrating the
power of certain methods of general application, and of
exhibiting a nice algebraic pattern: indeed, it turns out that
the problem is intimately connected to the classical problem of
decomposing the  \rep s of $sl(N)$  onto representations  of the
$B_l=so(2l+1)$ or $C_l =sp(2l)$ algebras, with $N=2l$ or $2l+1$.
This work generalises the previous results for $N=3$
\refs{\DIFZ,\BPPZ} and $N=4$ \AO.
As a  side result we derive various relations for the
$C_l^{(1)}$ modular matrices (see (2.11-15) below),
which to the best of our knowledge are new.

%%%%%%%%%%%%%%%%%%%%%%%%%%%%%%%%%%%%%%%%%%%%%%%%%%%%%%%%%%%%%%%

\subsec{The $A_{N-1}=sl(N)$ and the  affine $\slh(N)$ algebras}
\nind
To proceed, we need to introduce notations. As we are dealing
with pairs of Lie algebras, we consistently  use different types
of labels for their representations etc.  For the $\slh(N)$
theories under study, weights will be denoted by Greek letters.
At a given level $k$ or shifted level $\h=k+N$  these weights
belong to the Weyl alcove
\eqn\Aalc{ {\cal P}_{++}^{(A_{N-1},\,\h)}:=\{\Gl
=\sum_{i=1}^{N-1}\Gl_i \GL_i\ 
|\ 
\Gl_i\in \Bbb Z_{_{>0}}\, ,
 \sum_{i=1}^{N-1}\Gl_i \le \h-1\} \ ,}
where $\GL_i$, $i=1,\cdots,N-1$ are the $sl(N)$ fundamental
weights. The Weyl vector is $\rho=\sum_{i=1}\GL_i$.  The number
of weights in $\CP_{++}^{(A_{N-1},\,\h)}$ equals $ {\h-1 \choose
N-1}$. The alcove  is invariant under the action of $C$, the
complex conjugation of representations, $C\,:\
\Gl=(\Gl_1,\cdots,\Gl_{N-1})
\mapsto\Gl^*=(\Gl_{N-1},\cdots,\Gl_1)$, and of the $\Bbb{Z}_N$
automorphism $\sigma$, related to the cyclic symmetry of the
affine Dynkin diagram of type $A$
\eqn\sigmA{\sigma(\Gl)=(\h-\sum_{i=1}^{N-1}
\Gl_i,\Gl_1,\cdots,\Gl_{N-2})\ .}
Basic in our discussion is the symmetric, unitary matrix
$S=(S_{\Gl\Gm})$ of modular transformations. Under the action of
$C$ and $\sigma$,
\eqn\transfS{S_{\Gl^*\Gm}=S_{\Gl\Gm^*}=(S_{\Gl\Gm})^*\qquad 
S_{\sigma(\Gl)\Gm}=e^{{2\pi i \tau(\Gm)/ N}}\, S_{\Gl\Gm} \ ,}
where $\tau(\Gl):=\sum_{i=1}^{N-1} i(\Gl_i-1)$ is the $\Bbb{Z}_N$
grading of weights --the ``$N$-ality''.

We want to find a set of matrices 
$\{n_\zg\}_{\zg\in {\cal P}_{++}^{(N;\,\h)} }$ 
with non negative integer entries such that their matrix product reads
\eqn\Nimrep{ n_\Gl\, n_\Gm=\sum_{\Gn}\, N_{\Gl\Gm}{}^\Gn \, n_\Gn}
where $N_{\Gl\Gm}{}^\Gn$ are the fusion matrices of the $\slh(N)$ 
theory at that level. The $n_{\Gl}$ must satisfy Cardy
consistency condition 
\eqn\spec{
n_{\zl a}{}^b= 
\sum_{j\equiv j(\mu)\,,\,\, \mu \in\,  \Exp^{(\h)}}\, { S_{\zl \mu}\over
S_{\rho\mu}}\, \psi_a^j\, 
\psi_b^{j\,*}
}
with $\psi$ the unitary matrix diagonalising them; $j=j(\mu)$
labels a proper choice of basis.  Their eigenvalues are thus of
the form $\chi_{\Gl}(\Gm):=S_{\Gl\,\Gm}/S_{\Gr\,\Gm}$, 
and  are specified by the weights $\Gm$ labelling
the {\it diagonal} terms in  \modinv,  called ``exponents''.  In
the case under study, the exponents are the real, i.e.
self-conjugate, weights $ \Gm=\Gm^* $ in the alcove. Depending on
the parity of $N$, those have a different structure:
\eqn\realw{
\Exp^{\!(\h)}\ni \Gm=
\! \cases{ (m_1,\cdots,m_l,m_l,\cdots,m_1)\,, \
\ 2\sum_{i=1}^l m_i \le \h-1
& $\!\!\!\!\!\!\!\!\!\!\!\!$ if $N=2l+1$ 
\cr
(m_1,\cdots,m_{l-1},m_l,m_{l-1},\cdots,m_1)\,,
\ \ 2\sum_{i=1}^{l-1} m_i +m_l\le \h-1 \!\!\!\!& if  $\, N=2l$ 
\cr 
}}
and their number is 
\eqn\numbreal{
|\Exp^{\!(\h)}|={\rm\ \# real\ weights}= \cases{
{\lfloor{{\h-1\over 2}}\rfloor \choose l} 
 & if $N=2 l+1 $\cr  {\lfloor{{\h\over 2}}\rfloor\choose l}+ 
{\lfloor{{\h-1\over 2}}\rfloor\choose l  }    & if $N =2 l$ \cr }}

In general, the \nimrep\ matrices satisfy  $n_\Gl^T=n_{\Gl^*}$;
in the present case, because of the reality of the exponents
$\mu$, their eigenvalues $\chi_\Gl(\mu)$ are real and satisfy
$\chi_\Gl(\mu)=\chi_{\Gl^*}(\mu)$ and one concludes that the
matrices $n_\Gl$ are symmetric.  Moreover, because a real weight
$\Gm$ has a $N$-ality $\tau$ equal to $0\ \mod N$, resp. $0$ or
$N/2\ \mod N$, for $N$ odd, resp. even, eq. \transfS\ implies
that $n_\Gl$ is only a function of the orbit of $\Gl$ under
$\sigma$, resp. $\sigma^2$.  As usual, it is sufficient to find
the generators $n_{\GL_i+\rho}=n_{\GL_{N-i}+\rho}$ associated with the
fundamental weights to fully determine the \nimrep.  If the
matrices $n_\Gl=\big(n_{\Gl\,a}{}^b\big)$ are regarded as
adjacency matrices of graphs, it is natural to refer to the
labels $a,b$ of their entries as {\it vertices}. On the latter,
we do not know much a priori, besides that their number equals
the number of exponents
\numbreal. The set of vertices is denoted  by $\CV$.

Along with the \nimrep\ matrices $n_\Gl$, we are also interested
in finding a related set of matrices
$\hN_a=\big(\hN_{ba}{}^c\big)$, satisfying $\hN_a \hN_b=\sum_c
\hN_{ab}{}^c \hN_c$ and
\eqn\nNprod{n_\Gl\, \hN_a=\sum_{b\in \CV} n_{\Gl a}{}^b \, \hN_b \ .}
These matrices, associated with the vertices of the graph, span
the ``graph algebra'', which in this particular case is
commutative.  
The  set includes the unit matrix
attached to a special vertex denoted $1$ : $\hN_1=I$. Then the
previous relation evaluated for $a=1$ gives
\eqn\nNrel{n_\Gl=\sum_{b\in\CV} n_{\Gl 1}{}^b \, \hN_b\ ,}
i.e.  the \nimrep\ matrices are $\ge 0$ integer linear
combinations of the $\hN$.  The matrix $\psi$ in \spec\
diagonalises both $n$ and $\hN$ and \nNrel\ can be also rewritten
as
\eqn\char{\chi_\Gl(\mu)=\sum_{a\in\CV} n_{\Gl 1}{}^a\
\hchi_a(j(\mu))\,,\  \qquad \mu\in \Exp^{(h)}\,,
}
where $\hchi_a(j)=\psi_a^{j}/\psi_1^{j}$
are the eigenvalues of $\hN_a$.

In the present context the equations \nNprod, \nNrel\ have a
natural group theoretic interpretation. This is clear already in
the simplest case $N=3$ \BPPZ.  The reality of the exponents
\realw\ implies that they can be identified with an
integrable weight $\mu\to j(\mu)$ of $\slh(2)$ at a related
level. Then depending on the parity of $\h$, the coefficients $
n_{\Gl 1}{}^a$ originate from different patterns of decomposition
of the representations of  $sl(3)$ into those of $sl(2)$. Namely
the graphs are determined by the fundamental \nimrep\ which is
either $ n_{\zL_1+\rho \,1}{}^a=1+\delta_{a\, 2\Go}$, or $
n_{\zL_1+\rho \,1}{}^a=\delta_{a \, 3\Go}$, with $\Go$ the
$sl(2)$ fundamental weight, thus reflecting the two ways of
decomposing the $3$- dimensional $sl(3)$ representation.  As we
shall see, this example is the first in the series for odd $N$,
with  $C_l$ and $B_l$ taking over the r\^ole of $sl(2)$ for $\h$
even or odd respectively.  The ``branching coefficients''
interpretation of the \nimrep s and the equations \Nimrep,\nNrel\
has been discussed also in the  context of the discrete subgroups
of $SU(2)$.  See also \Qu\ for a related recent discussion.
Given the diagonalisation matrix $\psi_a^j$ one can compute as
well the structure constants of the algebra dual to the graph
algebra, the Pasquier algebra, which admits important physical
interpretations
\refs{\BPPZ}.

%%%%%%%%%%%%%%%%%%%%%%%%%%%%%%%%%%%%%%%%%%%%%%%%%%%%%%%%%%%%%%%%%%%%%%%%%

\subsec{$B$ and $C$ algebras}
\nind
We now briefly introduce relevant notations for the Lie algebras
$B_l$ and $C_l$  and their affine extensions $B_l^{(1)}$ and
$C_l^{(1)}$.

In the $B_l$ algebra, we denote the integrable weights by Latin
letters, keeping however the Greek $\Go_i$ for the fundamental
weights and $\bro$ - for the Weyl vector.  As the dual Coxeter
number  is $h^\vee=2l-1$, the  Weyl alcove at level $k$ is the
set
\eqn\alcB{{\cal P}_{++}^{(B_l,\h)} =\{m=\sum_{i=1}^{l} m_i\Go_i\ |\ 
m_i\in \Bbb Z_{_{>0}}\, ,
\ m_1+2 \sum_{i=2}^{l-1} m_i+m_l \le \h-1\}\ ,}
where the notation $\h$ is again used for the shifted level
$\h=k+2l-1$.  The number of integrable weights  is
$|\CP_{++}^{(B_l,\,\h)}|=  {\lfloor{{\h+1\over 2}}\rfloor\choose
l} + 2 {\lfloor{{\h\over 2}}\rfloor\choose l} +
{\lfloor{{\h-1\over 2}}\rfloor\choose l}\ . $ These weights are
graded  according to a ${\Bbb Z}_2$ grading $\tau(m):=  m_l-1
\mod 2 \ $ and the $\tau=0$ weights label  a subalgebra of the
Verlinde fusion algebra.  The ${\Bbb Z}_2$ automorphism of the
affine $B_l$ Dynkin diagram acts on  the weights in the alcove
as $\sigma(m)= (\h- m_1-2 \sum_{i=2}^{l-1} m_i-m_l\,,\, m_2,
\dots, m_l)$.
\medskip

For the $C_l$ algebra, we  use parallel notations: fundamental
weights are again denoted $\Go_i$, $i=1,\cdots,l$; the dual
Coxeter number is $h^\vee=l+1$ whence the shifted level
$\h=k+l+1$; the Weyl alcove reads
\eqn\alcC{  {\cal P}_{++}^{(C_l,\h)}=\{m=\sum_{i=1}^{l} m_i\Go_i\ |\ 
m_i\in \Bbb Z_{_{>0}}\, ,
\ \sum_{i=1}^{l} m_i \le \h-1\}\ ;}
the number of weights in the alcove is $| {\cal
P}_{++}^{(C_l,\h)}|={\h-1\choose l}$; the ${\Bbb Z}_2$ grading
reads $\tau(m):=  \sum_{i=1}^l i(m_i-1) \mod 2 \ $.  The ${\Bbb
Z}_2$ automorphism of the affine $C_l$ Dynkin diagram acts on
the weights in the alcove  as
$\sigma(m)=(m_{l-1},\dots,m_1\,,\,\h- \sum_{i=1}^{l} m_i)$.

The $S$ matrices of $B$ and $C$ type are real and satisfy a
$\Bbb{Z}_2$ analog of the $\sigma$ symmetry property \transfS.

%%%%%%%%%%%%%%%%%%%%%%%%%%%%%%%%%%%%%%%%%%%%%%%%%%%%%%%%%%%%%%%%%%%%%%%

\subsec{Results}
\nind
We may summarise our results as follows.  In general the
eigenvalues in the r.h.s.  of \char\ are expressed by  the
modular matrices $S$ of $B_l$ or  $C_l$
\eqn\heig{\hchi_a(j)={S_{aj}\over S_{1j}}\,,} in which the
weights of $B$ or $C$ algebras label both the graph vertices
$a\in \CV$ and the basis $j=j(\mu)$  into
\spec, related to a projection of the set of exponents \realw\
to the   $B$ or $C$  alcoves; the vertex $a=1$ in \heig\ is
identified with the $B$ or $C$ Weyl vector $\bro$, i.e.  the
shifted weight of the identity representation.  

\noindent
The situation depends on the parities of $N$ and of the shifted level $\h$.
\medskip
{\bf 1)} For $N=2l+1$, $\h$ even:
\medskip

\noindent
The set of exponents $\Exp^{\!(\h)}$ \realw\ is identified 
with the $C_l$ integrable alcove $\CP^{(C_l,{\h\over 2})}_{+,+}$
\eqn\ranexpc{
\CP^{(C_l,{\h\over 2})}_{+,+} \ni j(\mu)=(m_1,m_2,\dots,m_l)  \Leftrightarrow 
\mu=
(m_1,\dots,m_l, m_l,\dots,m_1)\in \Exp^{\!(\h)}
}
and the same alcove  parametrises  as well the set of graph
vertices $\CV\equiv \CP^{(C_l,{\h\over 2})}_{+,+} $.
 The Pasquier and  graph algebras are identical and coincide with the
$C_l$  Verlinde fusion algebra, $\hN_a=N_a$.  Accordingly
$\psi_a^j$  in \spec\ is provided by the $C_l$ modular matrix
$S$,
\eqn\psc{
 \psi_a^j =S_{aj} \,, \qquad a,j\in  \CP^{(C_l,{\h\over 2})}_{+,+}\,. 
}
The  decomposition formula \nNrel\ for the fundamental  \nimrep s  reads
\eqn\rel{
n_{\zL_i+\rho} =\sum_{k=0}^i\,  \hN_{\zo_{i-k}+\bro }\,,
\qquad i=1,2,\dots,l\,,
}
reproducing the  sl(3) result $n_{(2,1)}= I +\hN_{2\omega}$.
For $h=2l+2$ the alcove $\CP^{(C_l,{\h\over 2})}_{+,+}$ consists
of one point, the identity, and \rel\ degenerates to
$n_{\zL_i+\rho}= n_{\rho}=1$ for all $i$. 

\medskip
{\bf 2)} For $N=2l+1$, $\h$ odd:
\medskip

\noindent
The set of exponents  \realw\ is identified 
with a subset of the   alcove $\CP^{(B_l,\h)}_{+,+}$
\eqn\elmb{
{\rm Exp}^{(B)} 
=\{ \CP^{(B_l,\h)}_{+,+}\ni j= (m_1,m_2,\dots, m_l)\, |\, \tau(j)
=1 \,, \,  m_1< {\h-m_l\over 2} -\sum_{i=2}^{l-1}\, m_i\,\} \,,
}
where (note $\tau(j)+1=m_l=0$ mod $2$)
$$
{\rm Exp}^{(B)} \ni j=(m_1,m_2,\dots,m_l)  \Leftrightarrow 
\mu=
(m_1,\dots,{m_l\over 2}\,, {m_l\over 2},\dots,m_1)\in  \Exp^{\!(\h)}\,.
$$

\noindent
Another subset of $\CP^{(B_l,\h)}_{+,+}$ parametrises (for $l\ge 2$)
the  vertices
\eqn\alcb{
 \CV=\{ \CP^{(B_l,\h)}_{+,+}\ni a=(m_1,m_2,\dots, m_l)\,|\,  \tau(a)
 =0\,,\  m_1< {\h-m_l\over 2}-\sum_{i=2}^{l-1}\, m_i\,\}\,.
}
 For
$l=1$ the set of exponents and the set of vertices $\CV$ are
parametrised by the subsets of $\CP^{(A_1,h)}_{+,+}\,,\,$ $1\le m
\le {\h-1\over 2}\,,\,$
or $\,1\le m \le \h -2$, $m$ odd, respectively  \BPPZ.
The  eigenvector matrix in \spec\ is given for $l=1$ by
$\psi_a^j=\sqrt{2}\, S_{aj}$, and  for any  $l\ge 2$ by
\eqn\psb{
\psi_a^j=2\, S_{aj}\,, \quad a\in \CV\,, j\in  {\rm Exp}^{(B)} \,.
}
As empirically observed, there exists a basis (i.e., a preferred
correspondence of the two sets of indices  $\CV\,$,  ${\rm
Exp}^{(B)}$), in  which the matrix $\psi_a^j$ is symmetric and
hence the Pasquier algebra is identical to the  graph algebra.
The  matrices $\hN_a$  are expressed (for $l\ge 2$)   via the $B_l$ Verlinde
fusion  matrices,
\eqn\graver{
\hN_{ab}{}^c=N_{ab}{}^c-N_{ab}{}^{\sigma(c)}
\,, \qquad a,b,c \in \CV\,
}
and are  checked to be nonnegative. The fundamental
\nimrep s are 
\eqn\brel{
n_{\zL_i+\rho} =  \hN_{\zo_i+\bro}
,\ \ i=1,2,..., l-1\,, \qquad n_{\zL_l+\rho} = \hN_{2\zo_l +\bro}\,,
}
 reproducing for $l=1$
the sl(3) decomposition $n_{(2,1)}= \hN_{3\omega}$.
\medskip

{\bf 3)} For $N=2l$,  arbitrary $\h$:
\medskip
\noindent
The set of exponents \realw\ is identified with a subset of the 
alcove $ \CP^{(C_l,\h)}_{+,+}$
 \eqn\elmc{
 {\Exp^{(C)} }=\{\CP^{(C_l,\h)}_{+,+}\ni j= (m_1,m_2,\dots,
m_l)|\, \ m_1, \dots, m_{l-1} -{\rm even}\}\,,
}
$$
{\Exp}^{(C)} \ni j=(m_1,m_2,\dots,m_l)  \Leftrightarrow 
\mu=
({m_1\over 2},\dots,{m_{l-1}\over 2}\,, m_l\,,{m_{l-1}\over 2},\dots,
{m_1\over 2})\in  \Exp^{\!(\h)}\,.
$$
A subset of $\CP^{(C_l,\h)}_{+,+}$ parametrises the vertices
\eqn\alcc{
\CV= \CP^{(C_l,\lfloor{\h\over 2}\rfloor+1)}_{+,+}
 \cup \sigma_1(\CP^{(C_l,\lfloor{\h\over 2}\rfloor+1)}_{+,+})
\subset \ \CP^{(C_l,\h)}_{+,+} \ 
}
where
\eqn\sigf{
\sigma_1(m_1,\dots,m_l):=
(\h-m_1-2 \sum_{i=2}^l m_i, m_2, m_3, \dots, m_l)\,.
}
For  $\h$ odd \alcc\ is a disjoint union of two subsets of 
$\CP^{(C_l,\h)}_{+,+}$ of the same cardinality.
% equivalently 
% $$ \CV=\{\CP^{(C_l,\h)}_{+,+}\ni m= (m_1,m_2,\dots, m_l) 
% | m_1+2 \sum_{i=2}^l\, m_i \le h-1\}\,.$$

\noindent
The eigenvector matrix $\psi_a^j$ is expressed by the $C_l$ modular matrix
 $S$
\eqn\psc{
\psi_a^j =
(\sqrt{2})^{l-1}\, S_{aj}\,, \qquad a\in \CV\,,\  j\in \Exp^{(C)}\,.
}
Empirical data suggest that in general ($l>2$) $\psi$ in \psc\ is
not symmetrisable for $h$ even.  For $h$  odd $\hN_a$ are
nonnegative, while for $h$  even they may have signs. The same
applies to the matrices of the Pasquier algebra, in which the
role of the identity is played by $j=j(\rho)=(2,\dots, 2,1)$,
with $\psi_a^{j(\rho)}>0 $.  The fundamental \nimrep s are
\eqn\crel{
n_{\zL_i+\rho}=\sum_{m=0}^{\lfloor i/ 2\rfloor}\, 
\hN_{\zo_{i-2m}+\bro}\,, \qquad
i=1,2,\dots,l\,.
}
For $h=2l+1$ \crel\ degenerates to $n_{\zL_i+\rho}=\hN_{\bro}\,,$
for  $i$ even, $n_{\zL_i+\rho}=\hN_{\zo_1}\,,$ for $i$ odd, and
the graph algebra is isomorphic to $\Bbb Z_2$.  In general the
graph algebra matrices are expressed by the $C_l$ Verlinde
matrices
\eqn\graverc{
\hN_{ab}{}^c=\sum_{p=0}^{l-1}\,
\sum_{l\ge i_1 > i_2 > \dots > i_p \ge 2}\,
 (-1)^{\lfloor{i_1\over
2}\rfloor+\dots +\lfloor{i_p\over 2}\rfloor}\,
N_{ab}{}^{\zs_{i_1}\dots \zs_{i_p}\,\zs_1^{i_1+\dots +i_p}(c)}\,,
\qquad a,b,c \in \CV\,,
}
where $\zs_l=\zs\,,\, $ $\zs_1$ appears in \alcc, while $\zs_s$
for $s=2,\dots,l-1$ are defined below, see section 2.1 for more
details and the proof of \graverc.  

In the simplest example in this series, the case $\slh(4)$, the
formula \graverc\ reduces to two terms as in \graver, and  \crel\
reproduces the graphs displayed in \AO, see the Figure. In this
particular case $\CV$ is represented by the lower
``half-alcove'', i.e., the points $m=(m_1,m_2)\in
\CP^{(C_2,\h)}_{+,+}\,,$  $2 m_2< h-m_1$.

\medskip
\vbox{\centerline{\epsfxsize=14cm\epsfbox{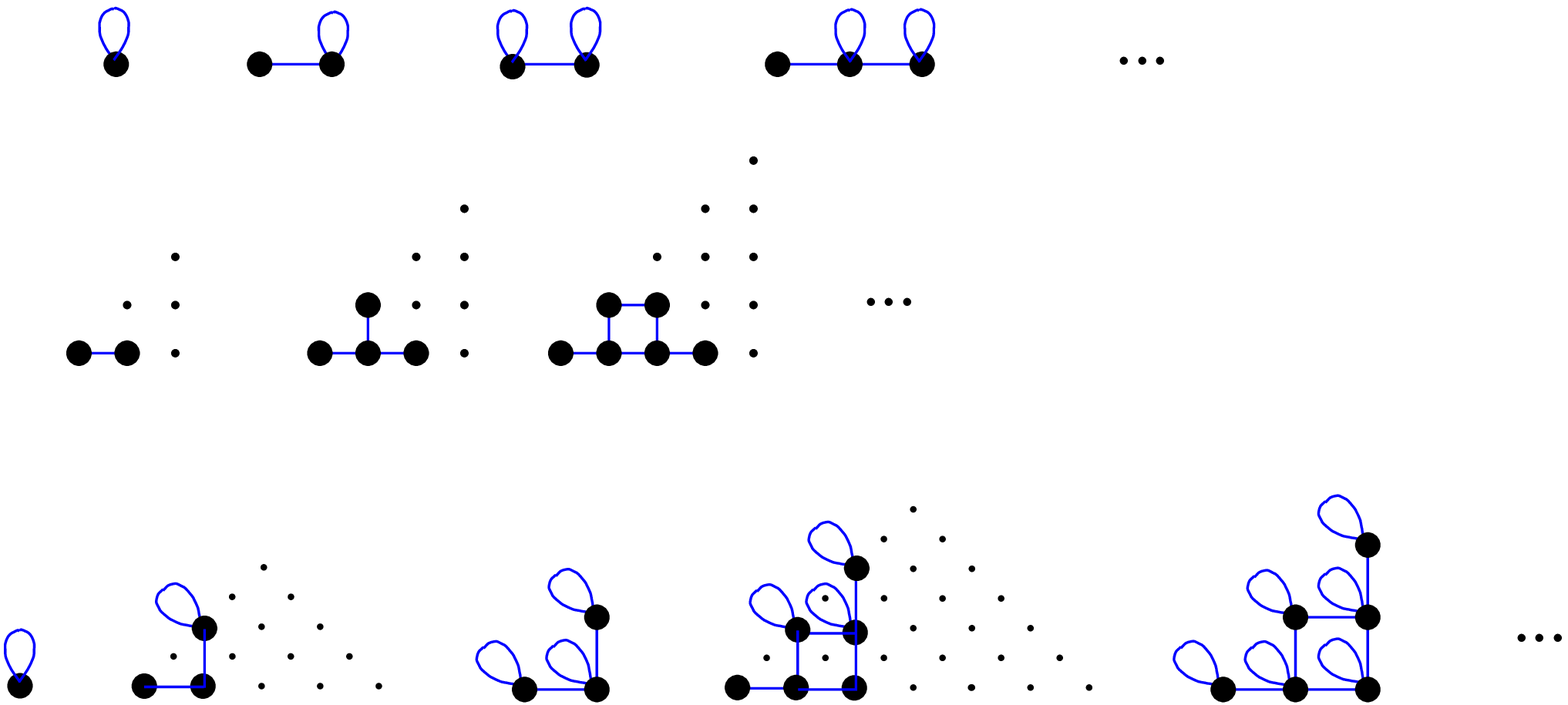}}
\medskip
{\bf Figure. } The graphs associated with the \nimrep\ matrices 
$n_{\GL_1+\rho}$ for $sl(N)$, $N=3,4,5$ and $h=N+1,N+2,\cdots$, 
drawn on the alcoves of $B$ or $C$ type.}

\bigskip

The formulae above ensure that the solutions of \Nimrep\ we find
are integer valued. Furthermore they are non-negative, however we
lack a general proof of this. 

Note that the  relations  \rel, \brel, \crel\ for $\h-N>1$
reflect three different  decompositions of the $A_{N-1}$
fundamental representations, in particular   the same equalities
hold  for the corresponding classical dimensions.

The rest of this note will review the two routes which led us 
to these results.
%%%%%%%%%%%%%%%%%%%%%%%%%%%%%%%%%%%%%%%%%%%%%%%%%%%%%%%%%%%%%%%%%%%%%%%%%%

\newsec{Decompositions of irreps -- classical consideration and
quantisation}
\nind 
The reality condition on the exponents in \realw\ leads to a
$\Bbb Z_2$ folding of the integrable alcove, or effectively, of
the set of $A_{N-1}$ roots, thus suggesting the relevance of the
algebras of $B$- or $C$- type. Recall that the Verlinde matrix
eigenvalues, or, fusion algebra characters, can be interpreted as
classical characters, evaluated on a discrete subset of the
Cartan group. The reduction of groups and their finite
dimensional irreducible representations (``irreps'') can be
described in terms of the weight diagrams $\CG_{\zl}$ associated
with the highest weights $\zl$ (in this ``classical'' discussion,
unshifted), or, in other words, as a decomposition
\eqn\chard{ 
\chi_{\zl+\rho}  = \sum_a\,  \tn_{\zl+\rho\, 1}{}^a \ \hchi_a\,.}
Here enter the formal characters, i.e., sums of formal
exponentials
\def\tu{\gamma}
$\chi_{\zl+\rho}=\sum_{\tu\in\CG_{\zl}}\,m_{\tu}^{(\zl)}\,
e^{\tu}\,,$ where $m_{\tu}^{(\zl)}\in \Bbb Z_{\ge 0}$ is the
multiplicity of the weight $\tu$ in the weight diagram
$\CG_{\zl}$ of the given irrep of highest weight $\zl$.  Thus the
first step in solving  the problem is the determination of the
classical decomposition formulae, i.e., the coefficients in
\chard, (whence the similar  notation as in \char), by choosing a
projection $P$ of the weights of $A_{N-1}$ and decomposing the
weight diagrams $\CG^{(A_{N-1})}_{\zl}$; in  the formal
exponentials $e^{\tu} \to e^{P(\tu)}$.  In the $A_2$ example
choosing $\tu\to P_1(\tu)=\tu_1+\tu_2$, or $\tu\to
P_2(\tu)=2(\tu_1+\tu_2)$, leads to
$P_1(\CG_{(1,0)}^{(A_2)})=\CG_{0}^{(A_1)}+\CG_{\Go}^{(A_1)}$,
or $P_2(\CG_{(1,0)}^{(A_2)})=\CG_{2\Go}^{(A_1)}$.     
It is sufficient to establish these relations for the fundamental
\rep s of $sl(N)$; to recover the remaining ones, we can use the
standard Pieri formulae.  The second step is based on the
alternative interpretation of the classical characters as group
characters, $\chi(H)$, evaluated on the Cartan subgroup $H$,
common to $SL(N)$ and its subgroup  $SO(2l+1)$ or $SP(2l)$.
Reducing to a subgroup means that typically the symmetry of the
formal characters is enlarged, e.g., we can identify the weight
diagram of a representation and its complex conjugate, and thus
in particular restrict to half the fundamental $A_{N-1}$
characters. The quantisation procedure leading to the fusion
characters  amounts to  evaluating the formal exponentials on
rational points $e^{\zg}({2\pi i\mu\over \h})=e^{2\pi
i\,\langle\zg,\mu\rangle\over \h}\,,$ $\mu \in \CP^{(\h)}_{+,+}$.
This selects  an embedding of the  range $\Exp^{\!(\h)}$ of
exponents in an integrable alcove of the smaller algebra.
E.g., for $A_{2}$, $\zl\in \CG_{(1,0)}^{(A_2)}$ and
$\mu=(m,m)\in \CP^{(A_2,\h)}_{+,+}$  one has $\langle
\zl,\mu\rangle^{(A_2)}=2 \langle P_1(\zl),m
\zo\rangle^{(A_1)}=\langle P_2(\zl),m\zo\rangle^{(A_1)}\,$, which
allows to identify   $\Exp^{(\h)}$  with the $A_1$ alcove
$\CP^{(A_1,{\h\over 2})}_{+,+}$  for $\h$ even, or with a  factor
$\CP^{(A_1,\h)}_{+,+}/\Bbb Z_2$  for odd $\h$.

Performing these two steps for the fundamental weights
$\zl=\zL_i\,,$ for which one can compute explicitly the
decomposition coefficients in  \chard, leads to the eigenvalue
decomposition formula \char\ for these weights.  In general the
``quantisation'' of the classical characters is equivalent to
enlarging further their symmetry, from the corresponding Weyl
$\bW$ to the affine Weyl group $W$, $\hchi_{w(a)}={\rm det}(w)\,
\hchi_a$, where $w(a)$ is an abbreviation for  the horizontal
projection of the action of $W$.  (Recall that we work with
shifted weights.) The symmetry allows to split the sum in \chard\
into a sum over weights $a$ in the corresponding integrable
alcove, and a sum over elements of $W$ which bring into the
alcove the (integral dominant) weights remaining outside of it.
In the case 1) this suggests the relation for  the coefficients
in \char\ in terms of the classical decomposition coefficients
 \eqn\quant{
n_{\zl 1}{}^a=\sum_{w\in W}\, {\rm det}(w)\, \tn_{\zl 1}{}^{w(a)}\,.
}
This relation is  analogous to a formula  \KWFGP\ expressing  the
Verlinde multiplicities via the classical multiplicities. In the
remaining two  cases presented in section 1.3.  the characters
$\hchi_a$ are evaluated on a smaller set than the integrable
alcove and the alcove has to be further factorised to a smaller
``fundamental'' domain, the set $\CV$. This effectively extends
the symmetry of the characters $\hchi_a$, and accordingly the
range of summation in \quant.
We postpone the discussion of the precise meaning of \quant\ to
the end of this section since in what follows we shall restrict
to fundamental weights. For $\zl=\zL_i$ the coefficients in
\char\ precisely coincide with their classical analogs in \chard\
in all but the smallest value of $\h$ in the cases 1) and 3) in
section 1.3.  E.g., in the $A_2$ example,
$\chi_{\zL_1+\rho}((m,m))=1+\hchi_{2\Go}(m)$, $m\zo\in
\CP^{(A_1,{\h\over 2})}_{+,+}$ for even $\h>4$, while for $\h=4$
the weight $2\Go$ is on the  $\bro$-shifted ``wall'' $m={\h\over
2}$ of the alcove  $ \CP^{(A_1,{\h\over 2})}_{+,+}$ (a fixed
point of the affine root reflection $w_0$), and hence the
character $\hchi_{2\Go}$ and the multiplicity $n_{\zL_1+\rho
\,1}{}^{2\Go}$ vanish.

Finally, finding an appropriate range $\CV$  of the same
cardinality as $\Exp^{\!(\h)}$, and an  unitary  matrix $\psi$,
allows to recover the matrix equivalent \nNrel\ of \char.

%%%%%%%%%%%%%%%%%%%%%%%%%%%%%%%%%%%%%%%%%%%%%%%%%%%%%%%%%%%%%%%%%%%%%%%%%%%

\subsec{Details and elements of proofs }

\nind 
Let  $\mu$ be a  weight in the weight lattice of $sl(N)$. 
Define the projections to weights of $B_l$ or $C_l$
\eqna\Pc
$$\eqalignno{
 \mu \to P_1(\mu) &=
(\mu_1+\mu_{2l}, \mu_2+\mu_{2l-1},\dots, \mu_l+\mu_{l+1})^{(C_l)} =
\sum_{m=1}^l \, 
\langle P_1(\mu)\,,\, \zb_{m}^{\vee}\rangle^{(C_{l})}\, \Go_m\cr
&=\sum_{m=1}^l \, \langle \mu\,,\,
\za_m+\za_{2l+1-m}\rangle^{(A_{2l})}\, \Go_m 
&\Pc {}\cr 
\Rightarrow  P_1(\zL_i)&=\Go_i\,,\, i=1,\dots ,l\,,\quad
  P_1(\za_i) ={\zb_i^{\vee}\over 2}\,,
}$$
\eqna\Pb
$$\eqalignno{ \mu \to 
P_2(\mu)&=(\mu_1+\mu_{2l},
\mu_2+\mu_{2l-1},\dots,2 (\mu_l+\mu_{l+1}))^{(B_l)} =
\sum_{m=1}^l \, \langle P_2(\mu)\,,\,
\zb_{m}^{\vee}\rangle^{(B_{l})}\, \Go_m\cr
&= \sum_{m=1}^{l-1} \, \langle \mu\,,\,
\za_m+\za_{2l+1-m}\rangle^{(A_{2l})}\, \Go_m+
2 \langle \mu\,,\, \za_l+\za_{l+1}\rangle^{(A_{2l})}\, \Go_l 
&\Pb {}\cr
\Rightarrow  P_2(\zL_i)&=\Go_i=\zo_i^{\vee}\,,\, i=1,\dots \,,
l-1\,,\  P_2(\zL_l)=2\Go_l=\zo_l^{\vee} \,, \quad
P_2(\za_i)=\zb_i\,, 
}$$
\eqna\Pcc
$$\eqalignno{
\mu \to P_3(\mu)&= (\mu_1+\mu_{2l-1}, \mu_2+\mu_{2l-2},\dots,
\mu_{l})^{(C_l)} =\sum_{m=1}^l \, \langle P_3(\mu)\,,\,
\zb_{m}^{\vee}\rangle^{(C_{l})}\, \Go_m\cr
&=\sum_{m=1}^{l-1} \, \langle \mu\,,\,
\za_m+\za_{2l+1-m}\rangle^{(A_{2l-1})}\, \Go_m+
 \langle \mu\,,\, \za_l\rangle^{(A_{2l-1})}\, \Go_l 
 &\Pcc {}\cr
 \Rightarrow  P_3(\zL_i)&=\Go_i\,,\, i=1,\dots ,l\,, \quad
  P_3(\za_i)=\zb_i\,.
}$$
Here ($\beta_i^{\vee}$) $\beta_i$  stand for the (dual) roots of
$B_l$ or $C_l$.  Clearly  $P_m(\zL_i)=P_m(\zL_{N-i})$. For
$l=1$ $C_1$ in \Pc{}\ and $B_1$ in \Pb{}\
are identified with $A_1$.   With the
exception of $P_2(\zL_l)=2\Go_l=\zo_l^{\vee}$
 (so that $\tau(P_2(\zL_i)) =0\
\forall i=1,\dots,l$) the maps $P_i$  relate the highest weights
of the fundamental representations of the pairs of algebras.
Applying these maps on the weights  
in the weight diagrams $ \CG_{\Lambda_i}$ of the fundamental
irreps of $A_{N-1}$ one  finds the coefficients
$\tn_{\zL_i+\rho \,1}{}^a$  for $\zl=\zL_i$ in \chard\
for each of the three cases:
\medskip
{\bf Proposition}: (Classical decomposition formulae)
\eqn\deca{
P_1(\CG_{\Lambda_i}^{(A_{2l})})=\oplus_{p=0}^i\, \CG_{\zo_{i-p}}^{(C_l)}\,,
}
\eqn\decb{  
P_2(\CG_{\Lambda_i}^{(A_{2l})})=\CG_{\zo_i^{\vee}}^{(B_l)}\,,
}
\eqn\decc{
P_3(\CG_{\Lambda_i}^{(A_{2l-1})})=\oplus_{p=0}^{\lfloor i/2\rfloor}\,
\CG_{\zo_{i-2p}}^{(C_l)} \,.
}

The next step is to interpret these decomposition formulae as
relations for the evaluated characters. The characters in the
l.h.s.  are evaluated on the exponent  subset $\Exp^{\!(\h)}$.
One finds that for $\zl\in \CG_{\Lambda_i}^{(A_{2l})}$ and $\mu\in
\Exp^{\!(\h)}\subset \CP^{(A_{2l},\h)}_{+,+}$
\eqn\beq{
\eqalign{
\langle\zl, \mu\rangle^{A_{2l}}&=
2 \langle P_1(\zl), (\mu_1, \dots, \mu_l)\rangle^{C_l}=2 \langle
 P_1(\zl), j(\mu) \rangle^{C_l} \,, \qquad 
j(\mu)= {P_1(\mu)\over 2}\,,\cr
\langle\zl, \mu\rangle^{A_{2l}}&=
\langle P_2(\zl), (\mu_1, \dots,\mu_{l-1}, 2 \mu_l)\rangle^{B_l}=
 \langle P_2(\zl), j(\mu) \rangle^{B_l} \,, \qquad j(\mu)=
{P_2(\mu)\over 2}\,. 
}}
In the first case the integrability condition for $\mu\in
\Exp^{\!(\h)}$ is equivalent to the $C_l$ integrability condition
for $j(\mu)$ at shifted level $\lfloor{\h\over 2}\rfloor$. For
$\h$ even, taking into account the factor $2$, one arrives at
\ranexpc.  The classical decomposition formula \deca\ determines
the coefficients in \char\ for the eigenvalues (fusion characters) 
$\chi_{\zL_i}(\mu)$ of the fundamental \nimrep s.
 These characters generate the $C_l$ Verlinde
fusion algebra which is diagonalised by the modular matrix $S$
and thus we recover the case 1) of section 1.3.

\medskip

For $\h$ odd we use instead the second equality in  \beq\
and  \decb. The map
$P_2$ embeds the exponent set $\Exp^{\!(\h)}$ into a subset of
the $B_l$ integrable alcove $ \CP^{(B_l,\h)}_{+,+}$, the one
described in \elmb, i.e., a $\Bbb Z_2$ - factor of the $\tau=1$
subset.  The eigenvalues    $\{\chi_{\Go_i^{\vee}+\rho}(j)\,,\,
i=1,2\dots,l\}$, labelled by the fundamental coweights (all of
$\tau=0$), generate an algebra $\hN_a$ with $a$ in the range
\alcb. Using the standard properties of the modular matrix $S$ it
is easily derived that this subalgebra is diagonalised by the
matrix in \psb, indeed 
$$
\delta_{jj'}= \sum_m\, S_{m j}S_{ij'}=2 \sum_{\tau(m)=0}\, S_{m j}S_{mj'}
=4\sum_{a\in \CV}\, S_{a j}\, S_{aj'} \,, \qquad j,j'\in {\rm Exp^{(B)}}\,.
$$
In the first step we have used that there are no $\sigma$  fixed
points in ${\rm Exp^{(B)}}$, while in the second, that
$S_{\sigma(m)j}=-S_{mj}$ for $j\in {\rm Exp^{(B)}}$, in
particular $S_{m_0 j}=0$ for the fixed points  $m_0=\sigma(m_0)$.
To prove the expression \graver\ for the   structure constants we
proceed similarly as above but this time obtain the exponent
region ${\rm Exp^{(B)}}$ by  factorising the $B_l$ alcove over
the $\Bbb Z_2$ symmetry of the $\sigma$ automorphism.  Namely
using that $S_{a\sigma(j)}=S_{aj}$, for $a\in \CV$ (since
$\tau(a)=0$), we  split the summation in the Verlinde formula for
$N_{ab}{}^c$ with all  $a,b,c \in \CV$ according to 
$$ 
N_{ab}{}^c
= \Big( 2\sum_{m\,,\, \langle m\,,\,\alpha_1 \rangle < \langle
\sigma(m)\,,\,\alpha_1\rangle } 
+\sum_{m\,,\, \langle m\,,\,\alpha_1 \rangle = \langle
\sigma(m)\,,\,\alpha_1\rangle}\, \Big)\,
{S_{am}S_{bm}S_{cm}\over S_{1m}}  \,.
$$ 
Note that since $\h$ is odd only points $m$ with  $\tau(m)=0$ may
contribute to the second term. We apply this formula to the
difference $N_{ab}{}^c-N_{ab}{}^{\sigma(c)}$, using that
$S_{cm}-S_{\sigma(c)m}=S_{cm}(1- (-1)^{\tau(m)})=0$ for
$\tau(m)=0$, while it gives a factor $2$ for $m\in {\rm
Exp^{(B)}}$, and hence we get \graver. This  reproduces the case
2) of section 1.3.

\medskip

Finally  one finds  for $\zl\in \CG_{\Lambda_i}^{(A_{2l-1})}$
and $\mu\in \Exp^{\!(\h)}\subset \CP^{(A_{2l-1},\h)}_{+,+}$
\eqn\ceq{
\langle\zl, \mu\rangle^{A_{2l-1}}= \langle P_3(\zl), 
(2\mu_1, 2\mu_2, \dots, \mu_l)\rangle^{C_l}= \langle
 P_3(\zl), j(\mu) \rangle^{C_l}  \,, \qquad j(\mu)=P_3(\mu)\,.
}
This determines the range \elmc\ for which the equality of
eigenvalues  implied by the classical formula \decc\ holds.
However the cardinality of this range, cf.  \numbreal, is
considerably smaller than the  cardinality of the full alcove  $
\CP^{(C_l,\h)}_{+,+}$ and this makes the last case 3) more
non-trivial.  The set $\CV$ in \alcc\  is obtained by a
$(l-1)$-step folding of the $C_l$ alcove $ \CP^{(C_l,h)}_{+,+}$.
Define recursively a sequence of involutive maps and a sequence
of subsets of the $C_l$ alcove
\eqna\seqsig
$$\eqalignno{
\sigma_s(m)&=(m_{s-1}, \dots, m_1\,,\,
\h-\sum_{k=1}^{s}\, m_k -2 \sum_{k=s+1}^{l}\, m_k\,,\,
 m_{s+1}, \dots m_{l})\,,\ m\in \CA_{s+1}  &\seqsig {}\cr
\CA_s&=\{ m\in \CA_{s+1}| \  m_s=\langle m,\alpha_s^\vee\rangle
\, <\, \langle  \sigma_s(m),
\alpha_s^{\vee}\rangle\ \}\,, \quad s=l,\dots, 1 \,,\ \CA_{l+1}=
\CP^{(C_l,h)}_{+,+} 
}$$
so that $\sigma_l=\sigma$. The set \alcc\ is easily seen to be
$\CV=\CA_2$ by comparing the inequalities that define both.
The map $\sigma_s$ can be interpreted as the horizontal
projection of the action of the elements $w^{(s)}=t_{\Go_s}\,
\bar{w}^{(s)}$ where $t_{\Go_s}$ is an affine translation in the
weight lattice, while  $ \bar{w}^{(s)}$ is an element in the
$C_l$ Weyl group $\bW\,,$  $\bar{w}^{(s)}: (e_1,e_2,\dots, e_l)
\to   (-e_s,-e_{s-1},\dots,-e_1, e_{s+1},\dots,  e_l)$. It keeps
invariant the subset  of  roots
$\{\za_1,\za_2,\dots,\za_l\,,\,-\za^{(s)} \}\,,\, $
$\za^{(s)}=e_1 +e_{s+1}$; $e_i$ are the orthogonal vectors
$\langle e_i,e_j\rangle={1\over 2}\, \delta_{ij}$,
$\Go_j=\sum_{i=1}^j\, e_i$, and by convention $e_{l+1}:=e_1$.
The transformations $\sigma_s$ lead to symmetries of the $C_l$
modular matrix, extending the analogous relation for
$\sigma=\sigma_l$:
\eqn\symS{
S_{\sigma_s(a)\,j}= (-1)^{s(j_l -1) +\lfloor s/2\rfloor  }\,
 S_{a\,j}\,, \quad  a\in 
\CA_{s+1}\,,  \ j\in
\Exp^{(C)}\,.
}
Indeed  the sign $ (-1)^{s+ \lfloor s/2\rfloor  }$ comes,  after
reordering in the Kac -Peterson formula for $S$, from the parity of   $
\bar{w}^{(s)}$, while for $j\in \Exp^{(C)}$ we have $e^{2\pi i
\langle\Go_s,w(j)\rangle}=e^{2\pi i j_l
\langle\Go_s, w(\sum_{i=1}^l e_i)\rangle}=(-1)^{sj_l}$ for any $w\in \bW$.
The relation \symS\ implies that  $\hchi_{\sigma_s(1)}(j)$ are the
eigenvalues of a simple current, i.e.  $\hchi_{\sigma_s(1)}(j)\,
\hchi_{ a}(j)=\hchi_{\sigma_s(a)}(j)$, and
$\{\hchi_{\sigma_s(1)}(j)\,,\, s=1,2,\dots,l\}$ 
generate  a group isomorphic to $(\Bbb Z_2)^{l}$.
\medskip

Furthermore  a relation stronger than \symS\  holds
for the particular points $m\in \CA_{s+1}$  satisfying
$m_s=(\sigma_s(m))_s=h-\sum_{p=1}^s m_p -2\sum_{p=s+1}^l m_p\,,\,$
for $1\le s\le l$, namely
\eqn\zerS{
S_{mj}=0\,, \qquad 
{\rm if}\quad 
\sum_{p=1}^l m_p=h-\sum_{p=s}^l m_p\,, \      j\in \Exp^{(C)}\,.
}
Proof: If $m$ has the property in \zerS\ then $w^{(1,s)}(m)=m-
\h(e_1+e_s)$ where $w^{(1,s)}\in \bW$ is the Weyl reflection
which maps $(e_1,e_s) \to (-e_s,-e_1)$ (for $s=1$, $e_1\to-e_1$),
keeping all the remaining $e_i$ unchanged. The Weyl group $\bW$
splits into pairs $\{w, w^{(1,s)} w\}\in \bW/\Bbb Z_2$ and the
contribution to $S_{mj}$ of each pair is zero.

The relations \symS, \zerS\ imply that the matrix defined in
\psc\ is unitary.  Indeed for $j+j'=0$  mod $2$ we have
$1+(-1)^{s(j_l+j'_l)}=2$ and using \symS\ and \zerS\ one proves
recursively that
$$
\sum_{a\in {\cal V}} \psi_a^j\psi_a^{j'}=
2^{l-2} \sum_{m\in \CA_{3}} S_{mj} S_{mj'}=\dots=\sum_{m\in
\CA_{l+1}} S_{mj} S_{mj'} =\delta_{jj'}\,.
$$
If instead $j_l+j'_l=1$ mod 2, hence $j\ne j'$,  one may 
write $ \sum_{a\in {\cal V}} \psi_a^j\psi_a^{j'}= \sum_{a\in
{\cal A}_1} \psi_a^j\psi_a^{j'} (1+(-1)^{j_l+j'_l})=0\  $ using
\symS\ and (for $\h$ even) \zerS\ for $s=1$.

Finally the formula \graverc\ is a consequence of the following
identity satisfied by the $C_l$ modular matrices $S$ with $c\in
\CV$ and $m\in \CP^{(C_l,\h)}_{+,+}$,
\eqn\alter{
\sum_{p=0}^{l-1}\,
\sum_{l\ge i_1 > i_2 > \dots > i_p\ge 2}\,
 (-1)^{\lfloor{i_1\over
2}\rfloor+\dots +\lfloor{i_p\over 2}\rfloor}\,
\,S_{\zs_{i_1}\dots \zs_{i_p}\,\zs_1^{i_1+\dots +i_p}(c)\, m}
=\prod_{i=1}^{l-1}\, \Big( 1+(-1)^{m_i}\Big)\ S_{cm}\ .
}
The r.h.s. of \alter\ is nonzero iff all $m_i=\langle m\,,\,
\alpha_i^{\vee} \rangle\,, \, i=1,\dots, l-1\,,$ are even, i.e.,
$m=j \in \Exp^{(C)}\,,$ in which case it equals $2^{l-1}\, S_{cj}$,
thus proving \graverc. The sign in \alter\ accounts for the
parity of the sequence of Weyl elements $\bw^{(s)}$ defined
above, $ (-1)^{\lfloor{i_1\over 2}\rfloor+\dots +\lfloor{i_p\over
2}\rfloor}= {\rm det}(\bw^{(i_1)}\dots \bw^{(i_p)}
(\bw^{(1)})^{i_1+\dots +i_p})$.  In particular the involution
$\zs_1$, which keeps invariant the subset $\CV$, contributes iff
$ i_1+\dots +i_p=1$ mod $2$.

We now sketch the proof of  \alter. As before the signs in
\alter\ are absorbed in the sum over $\bW$ defining $S$, while
using that $2 \langle 2 e_i\,,\, w(m)\rangle=0$ mod $2$, the sum
in the l.h.s. is  brought into the form
\eqna\proge
$$\eqalignno{
&S_{c\,m}+
{i^{l^2}\over \sqrt{(2\h)^l}}\,
\sum_{w \in \bW}\, \sum_{t=1}^{[{l\over 2}]}\,
 \Big(\sum_{j_1<j_2< \dots < j_{2t}}\, 
e^{2\pi i \langle e_{j_1}+\dots e_{j_{2t}}\,,\, w(m)\rangle} \Big)\,
{\rm det}(w)\, e^{-{2\pi i\over h} \langle c\,,\, w(m)\rangle}  \cr
&=\Big(1 +\sum_{t=1}^{[{l\over 2}]}\,  \sum_{j_1<j_2< \dots < j_{2t}}\,
e^{2\pi i \langle e_{j_1}+\dots e_{j_{2t}}\,,\, m\rangle} \Big)\,
S_{c\,m}\,. &\proge {}}$$  
The fact that only an even number of $e_i$'s appears is due to
the presence of $\zs_1$ in \alter.  In the last step we have used
that the sum of terms with fixed $t$ is $\bW$ invariant. Then
\proge{}\ implies \alter.  This completes the discussion of the
case 3).

%%%%%%%%%%%%%%%%%%%%%%%%%%%%%%%%%%%%%%%%%%%%%%%%%%%%%%%%%%%%%%%%%%%%%%%%%%

\subsec{General formulae for the \nimrep s}

We are now in a position to establish and generalise the formula
\quant\  in the introduction of this section, which also allows
to make connection with the discussion in \Qu.

In each of the three cases the boundary set $\CV$ arises as some
fundamental domain.  In  the case 1) $\CV$ coincides with the
integrable alcove $\CP^{(C_l,{\h\over 2})}_{+,+}$. In the case 2)
$\CV$ is the fundamental domain in the $\tau=0$ subset of the
integrable alcove $\CP^{(B_l,\h)}_{+,+}$ with respect to the
$\Bbb Z_2$ group  $\tW_+=\{1, \sigma\}$, using also that the
characters $\hchi_m(j)$ evaluated on the set of exponents  $j\in
\Exp^{(B)}$ vanish for the fixed points $m=\sigma(m)$. In other
words $\CV$ is the fundamental domain with respect to the action
of the extended affine Weyl group $\tW=\tW_+ \ltimes W$
in the lattice generated by the coweights $\omega_i^{\vee}$.
Finally in the case 3) denote by $\Gamma$ the set  of maps
$\gamma_{i_1,\dots, i_p}:= \zs_{i_1}\dots
\zs_{i_p}\zs_1^{i_1+\dots i_p}$ in \alter, \proge{},
with   det$(\gamma_{i_1,\dots, i_p}): =   {\rm
det}(\bar{w}^{(i_1)} \dots \,\bar{w}^{(i_p)}
\, (\bar{w}^{(1)})^{i_1+\dots i_p})$.
Any $m\in \CP^{(C_l,\h)}_{+,+}$, modulo the points  with
vanishing $\hchi_m(j) =0$  in \zerS\ (a subset of  cardinality
$|\CV|\,|\Gamma|=|\CV|\,2^{l-1}$), is represented uniquely as
$m=\gamma(c)\,, \, c\in \CV\,,\,
\zg\in \Gamma$. 

Let us introduce the notation   $W_{(i,\h)}$ to apply to the
three cases $i=1,2,3$ defined by the three projections $P_i\, $:
$\ W_{(1,\h)}$ is the $C_l$ affine Weyl group $W$ of the case 1),
$W_{(2,\h)}$ is the $B_l$ extended  affine Weyl group $\tW$
of the  case 2) and $W_{(3,\h)}=\{W\,\gamma\,, \ \gamma \in
\Gamma \}$. \footnote{${}^{1)}$}{ The elements of  $\Gamma$
(having translation parts in the root lattice, cf. \proge)
generate a finite group $W_{\Gamma}$.  For $l=2$,
$W_{\Gamma}=\Gamma=\{1,\sigma\}$, and $W_{(3,h)}=W_{\Gamma}
\ltimes W$  is the extended affine Weyl group.}  In all the cases
the characters satisfy $\hchi_{w(c)}(j)= {\rm det}(w)\,
\hchi_c(j)\,, \  w\in W_{(i,\h)}\,, \, c\in \CV$.

With these data at hand one has for $a,b\in \CV$
\eqn\ws{
n_{\zl+\rho\, a}{}^b = \sum_{\gamma \in \CG_{\zl}}\, 
m^{(\zl)}_{\gamma}\, \sum_{w\in W_{(i,\h)}} {\rm det}(w)\,
 \delta_{w(b)-a\,,\, P_i(\gamma)}=\sum_{w \in W_{(i,\h)}} {\rm
det}(w)\, \tn_{\zl+\rho\, a}{}^{w(b)} \,.
}
The derivation of this formula\footnote{${}^{2)}$}{See 
\Qu\  for a related recent discussion. The argument here
generalises the derivation of the analogous formula \KWFGP\ for
the fusion multiplicities, recovered  formally by
identifying the two algebras and setting $P=$Id.} follows the
same steps as the derivation of the corresponding formula for its
classical counterpart $\tn_{\zl+\rho\, a}{}^b $, analogous to the
first equality in \ws, but with $W_{(i,\h)} $ and $\CV$ replaced
by $\bW$ and the shifted by $\bro$
dominant Weyl chamber 
${\cal P}_+$.  Namely, using  the
Weyl formula for the character $\hchi_a$,  one first shows that
the multiplication in \nNprod\ is equivalent to a shift by the
weight diagram  $\CG_{\zl}$,
\eqn\der{
\chi_{\zl+\rho}\ \hchi_a 
= \sum_{\gamma \in \CG_{\zl}}\,
m^{(\zl)}_{\gamma}\, \hchi_{P_i(\gamma)+a}\,.
}
To get \der\ one exploits the fact that for any $w\in \bW$ there
is an element $w^{(A)}$ of the Weyl group of $A_{N-1}$, s.t.
$P(w^{(A)}(\gamma))=w(P(\gamma))$, and  also uses   the $\bW^{(A)}$-
invariance of the classical multiplicities
$m^{(\zl)}_{w^{(A)}(\gamma)}= m^{(\zl)}_{\gamma}$ (recall that here
$\zl, \gamma$ are unshifted weights).
At  the last step, leading to \ws, one takes into account the
fundamental domain property of each $\CV$ and the symmetry
properties of the characters derived above.

Inserting  \ws\ for $a=\bro=1$ into   \nNrel\  (with $\zl$
denoting now a shifted weight) we can write also
\eqn\wsa{
n_{\zl} 
=\sum_{c\in \CV}\, \Big(\sum_{w \in W_{(i,\h)}} {\rm det}(w)\,
\tn_{\zl\, 1 }{}^{w(c)}\Big)\ \hN_c\,.
}
This is a generalisation of the formulae \rel, \brel, \crel\ for
the fundamental \nimrep s, in which the alternating sum in \wsa\
reduces to the classical branching coefficients $\tn_{\zl\,
1}{}^{c}$.  In the case 1) \wsa\ is a sum over the Verlinde
matrices $\hN_c=N_c$, while in the cases 2) and 3) we  can
furthermore replace $\hN_c$  in \wsa\ with the Verlinde
submatrices $N_{a c }^b\,, \, a,b\in \CV$, using \graver\ and
\graverc.  In the former case the resulting formula can be
brought into the same form as \wsa,  with the same range of $w$,
but with  $c\in \CP^{(B_l,\h)}_{+,+}\,,$ $\tau(c)=0$.

Using \ws\ one can also  show that in all three cases the set
of coefficients $n_{\zl 1}{}^1$, $\zl\in
\CP^{(A_{N-1},h)}_{+,+}$  are determined by the classical
multiplicity of the identity representation $\bro=1$
in the $A_{N-1}$ irrep of highest weight $\zl$, i.e., $n_{\zl
1}{}^1=\tn_{\zl 1}{}^1$.

%%%%%%%%%%%%%%%%%%%%%%%%%%%%%%%%%%%%%%%%%%%%%%%%%%%%%%%%%%%%%%%%%%%%%%%%%%%
\newsec{Xu algorithm}
\nind
In this section we sketch a general method  for the construction
of the \nimrep\ starting from a limited information. Although
this method has been proposed and used in a more abstract form
already a while ago \refs{\Xu,\BEK}, its algorithmic and
systematic character may not have  been stressed enough. In this
approach, the fundamental datum is the set of \rep s $\Gl$ which
give a non vanishing $n_{\Gl 1}{}^1$. The fact that these data
play a central r\^ole has been discussed before \refs{\BPPZ,\BEK}
and stressed also recently in the category theory approach to
these questions \refs{\KO,\FRS}.  The meaning of these data is
best understood in block diagonal theories, since there, the
vertex denoted 1 corresponds to the block of the identity \rep,
and  $n_{\Gl 1}{}^1$ thus tells us which \rep s appear in that
block.

One may now proceed in three steps\par
1. The data  $n_{\Gn 1}{}^1$ are subject to a constraint coming
from \spec\ and the unitarity of the $S$ matrix 
\eqn\constr{\sum_{\Gn\in \CP_{++}}\Big(\sum_{\Gm\in
\Exp^{\!(\h)}}S_{\Gl\Gm}S_{\Gn\Gm}^* 
-\delta_{\Gl\Gn} \Big)n_{\Gn 1}{}^1=0\ .}
It is thus helpful to have a good Ansatz for the particular
matrix element $n_{\Gn 1}{}^1$ and to check that it passes the
test of \constr.  As explained in sect 1.1, in the present
problem, $n_{\Gn 1}{}^1$ is only a function of the $C$- and
$\sigma^\#$-orbit of $\Gn$, with $\#=1,2$ for $N$ odd, even, and
this reduces greatly the number of unknowns.

2. One then regards the yet undetermined matrices $n_\Gl$ as
forming a  set of vectors and one makes use of the \nimrep\
property \Nimrep, taking its $1,1$ matrix element,  to define a
symmetric scalar product between these vectors according to
\eqn\scalpr{\langle n_\Gl, n_\Gm\rangle:= (n_\Gl.n_{\Gm^*})_1{}^1
=\sum_a n_{\Gl 1}{}^a n_{\Gm 1}{}^a
=\sum_{\Gn}
N_{\Gl\Gn}{}^\Gm  n_{\Gn 1}{}^1 \ .}
In particular the identity matrix $n_\Gr=I$ associated with the
trivial representation has norm 1.  To the Gram matrix of scalar
products one may then apply the Schmidt orthogonalisation method
to determine a basis of vectors, ${\widetilde{n}}_\Gl$,  which
are obtained from the original  $n_\Gl$ by a triangular integer
valued matrix, starting from $\tilde n_\rho=n_\rho=I$, and which
are mutually orthogonal. In the most favourable cases, such as
those presently discussed, all these orthogonal basis vectors
have a norm equal to 1. This means that $\sum_a
\big({\widetilde{n}}_{\Gl 1}{}^a\big)^2=1$, hence there exists a
unique vertex $a_\Gl$ contributing to this matrix element: $
{\widetilde{n}}_{\Gl 1}{}^a=\delta_{a \, a_\Gl}$.  (If the
squared norms  of some ${\widetilde{n}}$   are equal to 2 or 3,
one has rather a sum of two or three Kronecker deltas, while
squared norms larger or equal to 4 present new options that have
to be examined in turn: either some entries $
{\widetilde{n}}_{\Gl 1}{}^a$ are larger than 1, or there are more
vertices contributing to them, see \Xu\ for examples.) The above
triangular system induces  an order $\prec$ between the labels of
the basis and one may invert it to determine the original $n_\Gl$
as a  sum of such contributions
\eqn\det{n_{\Gl 1}{}^a= \delta_{a a_\Gl}+
\sum_{\Gm\prec \Gl} b_{\Gl\Gm} \delta_{a a_\Gm},\qquad 
b_{\Gl\Gm}=
n_{\Gl 1}{}^{a_\Gm}\in \Bbb{Z}_{\ge 0}\ .}
The set of vertices $\CV$ is made of the $a_\Gl$, 
where $\Gl$ labels  the basis $\widetilde{n}_\Gl$.

3. In the last step one reconstructs the whole \nimrep\ from the
knowledge of the entries $n_{\Gl 1}{}^a$. This is done
recursively.  Assume first for simplicity that all
$\widetilde{n}_\Gl$ in step 2 have norm 1.  Take for example any
fundamental \rep,  called here  \def\f{{\GL}}$\!\!$ $\f$  
generically. Suppose all entries $n_{\f a_\Gm}{}^a$ have been
determined for $\Gm\prec\Gl$ and evaluate in two different  ways
$(n_\Gl n_\f)_1{}^a= n_{\f a_\Gl }{}^a+ \sum_{\Gm\prec\Gl}
b_{\Gl\Gm} n_{\f a_\Gm }{}^a =\sum_\Gn N_{\Gl  \f}{}^\Gn
n_{\Gn\,1}{}^a$, a relation which determines $ n_{\f a_\Gl
}{}^a$.  If some  $\widetilde{n}_\Gl$ has a norm $>1$ and
$n_{\Gl1}{}^a=\sum_\alpha \delta_{a a_{\Gl_\alpha}}$, this leads
to dichotomic choices, as we have to split the sum $\sum_\alpha
n_{\f a_{\Gl_\alpha}}$ into individual matrices, but at any rate,
this is  a finite problem.

This procedure has been applied successfully to the problem at
hand. In step 1, one  finds again a dependence on the parity of
$N$ and/or of $\h$.  While for odd $N=2l+1$ and $\h$ even,
$n_{\Gl\, 1}{}^1=1 $ for all $\Gl$ in the alcove, for $\h$ odd,
$n_{\Gl\, 1}{}^1\ne 0 $ and in fact $=1$ iff
$\Gl=(\Gl_1,\cdots,\Gl_{2l})$ has all its Dynkin labels $\Gl_i$
odd; for even $N=2l$, the same happens iff
$\Gl=(1,\Gl_2,1,\Gl_4,\cdots,\Gl_{2l-2},1)$, i.e. when $\Gl$ has
Dynkin indices of odd rank equal to 1.  In step 2, one may find a
basis of orthonormal vectors, i.e. all norms are 1. Finding this
basis may be laborious, and some external information, like the
one coming from the embedding picture discussed in the previous
section, is helpful but  not necessary.  Last step 3 then
proceeds in a straightforward way, and the whole algorithm has
been implemented in Mathematica.

The algorithm amounts effectively to solving recursively the two
sets of equations \Nimrep, \nNprod.  To make contact with \KO\
note that the graph can be alternatively described identifying
the vertices with the matrices $X_a:=\sum_{\zl}\, n_{\zl 1}{}^a\,
n_{\zl}\,$,   which satisfy an equation analogous to \nNprod,
$n_{\zl} \,X_a=\sum_{b}\, n_{\zl a}{}^b\,\, X_b\,$, with
non-trivial $X_1$.  Giving a solution for the coefficients
$n_{\zl 1}{}^1$ determines $X_1$  or the algebra $A$ of \KO, and
the graph is recovered starting from   $X_1$  and using
recursively as above the system \Nimrep.

\bigskip
%%%%%%%%%%%%%%%%%%%%%%%%%%%%%%%%%%%%%%%%%%%%%%%%%%%%%%%%%%%%%%%%%%%%%%%%%%%
\noindent {\bf Acknowledgements}. 
\medskip
We are indebted to  V. Ostrik for a question which added to the
motivation for this work. Thanks also to T. Quella and V.
Schomerus for discussions and to T. Gannon for informing us on
the existence of their  related work in progress with M.
Gaberdiel, see \GG.
\medskip

Note added: A different approach to the problem of describing the
boundary conditions associated with \modinv\ has been developed
earlier in \BFS, we thank J. Fuchs and C.  Schweigert for
informing us about this work.  In it the  sets $\CV$ and $\Exp$
are parametrized by highest weights of integrable representations
of twisted affine algebras; more details and comparisons can be
now found in \GG. Let us note that for even $N$ the unitary
eigenvector matrix \psc\ given  in terms of particular
$C_l^{(1)}$ modular matrix elements coincides 
with the $S$ matrix relating the  characters of integrable
representations of the twisted affine algebras $A^{(2)}_{2l-1}$
and $ D^{(2)}_{l+1}$ (see the formula for the latter S matrix in
Theorem 13.9 in the book of V. Kac  in  \KWFGP\ ).
 
%%%%%%%%%%%%%%%%%%%%%%%%%%%%%%%%%%%%%%%%%%%%%%%%%%%%%%%%%%%%%%%%%%%%%%%%%%%

\bigskip

\listrefs
%%%%%%%%%%%%%%%%%%%%%%%%%%%%%%%%%%%%%%%%%%%%%%%%%%%%%%%%%%%%%%%%%%%%%%%%%%%
\bye